# Privacy Preserving Data Aggregation in Wireless Sensor Networks

Arijit Ukil
*Innovation Labs, Tata Consultancy Services, Kolkata, India*
arijit.ukil@tcs.com

**Abstract**

*Privacy preservation is an important issue in today's context of extreme penetration of internet and mobile technologies. It is more important in the case of Wireless Sensor Networks (WSNs) where collected data often requires in-network processing and collaborative computing. Researches in this area are mostly concentrated in applying data mining techniques to preserve the privacy content of the data. These techniques are mostly computationally expensive and not suitable for resource limited WSN nodes. In this paper, a scheme is developed to provide privacy preservation in a much simpler way with the help of a secure key management scheme and randomized data perturbation technique. We consider a scenario in which two or more parties owning confidential data need to share only for aggregation purpose to a third party, without revealing the content of the data. Through simulation results the efficacy of our scheme and compare the result with one of the established scheme [1].*

## 1. Introduction

The explosive growth of hardware, software along with immense computing and communication power of the system and devices, it is unbelievably easy to store, retrieve and process large amounts of information. Good amount of privacy issues also arise with the proliferation of digital technologies. Explosive progress in networking, storage, and processor technologies has led to the creation of ultra large databases that record unprecedented amount of transactional information. In tandem with this dramatic increase in digital data, concerns about informational privacy have emerged globally. Privacy issues are further exacerbated now that the World Wide Web (WWW) makes it easy for the new data to be automatically collected and added to databases [2]. With ubiquitous connectivity, people are increasingly using electronic technologies in business-to-consumer and business-to-business settings. This in effect helps a third party to acquire the confidential and private information from various avenues. Depending upon the nature of the information, users may not be willing to divulge the individual values of records. This has lead to concerns that the private data may be misused for a variety of purposes. Privacy can be defined as the limited access to a person or a process and to all the features related to the person or the process. Privacy preservation is important from both individual as well as organizational perspectives. For example, customers might send to a remote database queries that contain private information. Two competing commercial organizations might jointly invest in a project that must satisfy both organizations' private and valuable constraints, and so on. In order to alleviate these concerns, a number of techniques have recently been proposed to perform the data mining tasks in a privacy-preserving way, which is called Privacy Preserving Data Mining (PPDM). The research of PPDM is aimed at bridging the gap between collaborative data mining and data privacy. Privacy-preserving data mining finds numerous applications in surveillance, in-network processing which are naturally supposed to be "privacy-violating" applications. The key is to design methods [4] which continue to be effective, without compromising security.

In this paper, we consider a scenario where data aggregation needs to be done in privacy-preserved way for distributed computing platform. There are number of data sources which collect or produce data. The data collected or produced by the sources is private and the owner or the source does not like to reveal the content of the data. But the collected data from the source is to be aggregated by an aggregator, which may be a third party or part of the network, where the data sources belong. The data sources do not trust the aggregator and like to hide the content of the data. So the data needs to be secure and privacy protected. In tune of that, we propose a scheme which is secure and privacy preserved. The computation for the aggregation is based on the concept of Secure Multiparty Computation (SMC). Generally, this problem can be seen as a computation of a function $f(x_1, x_2, ..., x_N)$ on private inputs $x_1, x_2, ..., x_N$ in a distributed network with n participants where each participant i knows only its input $x_i$ and no more information except output $f(x_1, x_2, ..., x_N)$ is revealed to any participant in the computation. In our case the function is SUM. We apply the property of modular arithmetic to recover the aggregated value. In our scheme, privacy is preserved through randomization process. The security part is handled by random key pre-distribution method. The scheme is simple in nature with low computational complexity, which makes it suitable for practical implementation particularly in the case where the source nodes do not have much computational capabilities. We compare the results, importantly the computational complexity performance of our scheme with that of described in [1], which has the same objective of secure privacy-preserving data aggregation. The proposed scheme has two parts:
1. Secure key management
2. Privacy preservation

The paper is organized as follows. In next section, related works in the field of privacy preservation is highlighted. In section 3, the considered system model is shown. After that, in section 4, we present the key distribution scheme. We present privacy preservation scheme in section 5. In section 6, we describe and analyze the simulation results. Finally, we conclude the paper in section 7.

## 2. Related Work

In the literature, number of techniques has been illustrated to effectively preserve the privacy of the source data. One of most popular method is randomization. The randomization method is a technique in which noise is added to the data to be privacy-protected. This is done to mask the attribute values of records [5]. The noise added to the data is sufficiently large so that individual values cannot be recovered. Therefore, techniques are designed to derive aggregated distributions from the perturbed data values. Subsequently, data mining techniques can be developed in order to work with these aggregate distributions. The randomization method has been traditionally used in the context of distorting data by probability distribution for methods such as surveys. There are two major classes of privacy preservation schemes are applied. One is based on data perturbation techniques, where certain distribution is added to the private data. Given the distribution of the random perturbation, the aggregated result is recovered. In another technique, randomized data is used to data to mask the private values. However, data perturbation techniques have the drawback that they do not yield accurate aggregation results. It is noted by Kargupta et al. [6] that random matrices have predictable structures in the spectral domain. This predictability develops a random matrix-based spectral-filtering technique which retrieves original data from the dataset distorted by adding random values. There are two types data perturbation. In additive perturbation, randomized noise is added to the data values. The overall data distributions can be recovered from the randomized values. Another is multiplicative perturbation, where the random projection or random rotation techniques are used in order to perturb the values. In tune of their argument, we apply the second technique of masking the private data by some random numbers to form additive perturbation.

Our objective of privacy preserved secured data aggregation falls under the broad concept of Secure Multiparty Computation (SMC) [7-11]. SMC and privacy preservation are closely related, particularly when some processing or computation is required on the data records. Historically, the SMC problem was introduced by Yao [3] where a solution to the so-called Yao's Millionaire problem was proposed. In general SMC problem deals with computing any (probabilistic) function on any input, in a distributed network where each participant holds one of the inputs, ensuring independence of the inputs, correctness of the computation, and that no more information is revealed to a participant in the computation than can be inferred from that participant's input and output [7]. The aggregation methods of privacy-preservation are dealt well in [8]. In [1], Wenbo He et.al. propose schemes to achieve data aggregation while preserving privacy. The scheme they proposed, CPDA (Cluster-based Private Data Aggregation) performs privacy-preserving data aggregation in low communication overhead with high computational overhead.

In this work, we propose the privacy-preserving data aggregation scheme which has much lower complexity than the CPDA scheme in [1]. We have shown in our scheme that simple use of modular arithmetic and additive randomization of the data records can be sufficient for privacy preservation.

## 3. System Model

In this section we present the system model, based on which the scheme is developed. This is shown in fig. 1. It is shown that there are N numbers of source nodes or sources which collect or produce the private data. These sources are the owners of the private data. Against the query of the service provider or the server, the sources answer the query of the server. In this process, the sources never reveal the content of the private data, i.e. they never share the private data in raw form. They perform some data perturbation technique on the raw data, from which the server cannot understand the original content of the data. The function of the server is to aggregate the data received from n servers. It may send the aggregated data value for further processing. It is also assumed that for each source at least one source is connected. The aggrgator or server node has the responsibility of data aggregation and further processing of the aggregated data. This server node has connection with N number of source nodes, which are connected with the server node through wireless links. These source nodes collect the data on its own or as per the instruction by the server node. It is assumed that the also source nodes have peer-to-peer connectivity atleast with one of the nodes inorder to reach the aggregator. We define data aggregator function as:

$$Y(t) = f(d_1(t), d_2(t), \ldots, d_N(t))$$

Without loss of generality, we consider sum function, though typical aggregation functions like mean, min, max can also be included.

$$Y(t) = \sum_{n=1}^{N} d_n(t)$$

In order to model the individual behaviors of the participating parties, we adopt the semi-honest model [7] that is commonly used in SMC. A semi-honest party follows the rules of the protocol, but it can later use what it sees during execution of the protocol to compromise other parties' data privacy. Such kind of behavior is referred to as honest-but-curious behavior [7] or passive logging [9]. The semi-honest model is realistic for our context as each participating party will want to follow the agreed protocol to get the correct result for their mutual benefits and at the same time reduce the probability and the amount of information disclosure about their private data during the protocol execution due to competition or other purposes.

In order to counter both the problem of security and privacy, our scheme consists of a secure key distribution scheme along with the privacy preservation mechanism. In fact, our scheme has two separate parts. First one is the key distribution, which guarantees the secure communication. This key distribution is part of the security system. Other part is privacy preservation through modular arithmetic. This key distribution method is described next followed by our privacy preservation scheme.

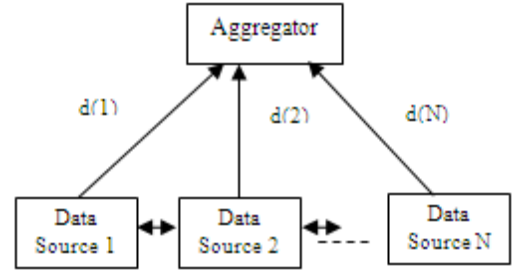

**Fig. 1 Privacy preserved data aggregation system model**

## 4. Secure Key Distribution

For secure key distribution, we follow the method proposed by Eschenauer and Gligor, which is called random key pre-distribution [10]. In our scheme we have K number of keys, which is stored in every source node. Of that K-k keys are shared with the server/aggregator for source to aggregator secure communication and k number of keys are kept for source to source communication. Our secure key distribution method has two parts.
1. Aggregator to source key exchange
2. Source to source key exchange

*4.a Aggregator to source key establishment:*

It is described that each source node has K-k number of keys shared with the server. As, all the source nodes possess the same keys, it is totally unsecure when a source node communicates with the server node with the shared key. Any malicious source node can decipher the source nodes' communication with the server and can launch attack very easily. In order to avoid this, in the pre-distribution phase, the source-aggregator key bank is randomly permuted and reordered for each source-aggregator pair. This ordering of the key bank is stored in the server for each source. When a new source node is added, same procedure of key bank randomization is followed and the key order is stored offline for the new source. Now, the source node communicates with the server through one of its shared keys. To accomplish this action, the source node first generates a random number between 1 and K-k. This random number (Rn) is sent to the server in plain text. The server understands that the source node will encrypt the next message by the Rnth number key of the key bank. Every time the source node likes to communicate with the server, it does the same steps. The random number generation process at each session hinders the probable guessing attack. It can also be observed that the random number (Rn) is sent to the source in plain text. But this does not arise any vulnerability issue, as getting the random number by a malicious node does not harm because of the randomization of the key bank order. The $Rn^{th}$ number key is different in different source nodes. The mapping is stored in the server offline in pre-distribution stage.

*4.b Source to source key establishment:*

In order to accomplish source to source key establishment, we assume that source to aggregator/server key is securely established. So, source to source key establishment is done through server only not directly between the sources. We observe that the k keys are same for all the source nodes, it becomes easy for another source node to decrypt the information of other sources, i.e. source 3 can decrypt what source node 1 and source node 2 are communicating. To avoid this situation, source node 1 and source node 2 separately permute the key bank order of the k number of keys dedicated for source-source communication and reorder that randomly. After that, they pass the permute function to each through the server using their pair-wise key with the server. After successful delivery of permute functions, one of the source nodes (source node 1, for example) sends another random number between 1 and k to the other source node (source node 2), which indicates the particular key of the permuted key bank. This pair-wise key between source nodes will be used for the subsequent communication until the data aggregation is complete. For next round of data aggregation process, same key establishment procedure will be followed.

## 5. Privacy Preservation

In our system model, there are N numbers of source nodes. Each source i owns a value $x_i$ which it is not willing to share with other parties. Suppose that the sum is in the range [0, M]. Our objective is to find out the sum X privately without revealing the private data $x_i$, i=1,2, … , N to each other as well as to the server.

$$X = \sum_{i=1}^{N} x_i$$

The process is initiated by the server. The server randomly chooses one of the source nodes and signals it to initiate the process. The source node first chosen by the server is denoted by $c_1$. This node possesses its private data $x_1$ and it generates one random number $r_1$ between the range [0, M], which is denoted as $r_1$ It then computes $R_1$.

$$R_1 = (r_1 + x_1) mod X$$

After computing $R_1$, the source node $c_1$ performs neighborhood discovery to find out the other source nodes it is connected to. This information $c_1$ passes to the server. Server keeps the knowledge of the nodes already participated. If the source nodes connected to $c_1$ is not already participated, the server randomly chooses one of those non-participated source nodes and sends that message to $c_1$. Let this next source node be $c_2$. Now, accordingly $c_1$ passes $R_1$ to $c_2$.

The source node $c_2$ computes $R_2$.

$$R_2 = (R_1 + x_2) mod X$$

The source node follows the same procedure as c1 and sends $R_2$ to $c_3$. This way $c_N$ is reached, which computes $R_N$.

$$R_N = (R_{N-1} + x_{N-1}) mod X$$

The server when finds out that all the nodes are participated, it asks the last node to send $R_N$ to it. Server now directs the first source node $c_1$ to compute the summation as:

$$X = (R_N - R_1) mod X = \sum_{i=1}^{N} x_i$$

The source node after computing the summation sends that value to the server. The server may process it or sends that value for further processing.

In the case the server/aggregator finds that the neighborhood nodes are all participated and still few source nodes are left and these source nodes are not with the neighborhood range of the last source (who does the latest computation and sends that message to the server), the server first randomly chooses one of the sources not participated. It then sends this message to the last participated source. This source then sends the computed value to the server. Server/aggregator forwards the computed value to the chosen node.

It can be observed that that each party is assumed to have used their correct value $x_i$. If there is no collusion, party i only learns the total sum x, and can also calculate $(x − x_i)$ mod n, i.e., the sum of values of all the other source nodes. However, if two or more source nodes collude, they can disclose more information. For example, if the two neighbors of party i (that is, parties $i − 1$ and $i + 1$) collude, they can learn $x_i = (R_i − R_{i-1})$ mod n. We have intelligently averted this possibility by allowing the server or aggregator to decide the next source node. Server does that by choosing randomly from the eligible neighbor of the source node. But there exists a possibility of colluding through bypassing the server. In that case, the source node sends the computed $R_i$ to its friend node (other colluding node), the scheme needs to be slightly modified. In that scenario, source to source communication needs to be strictly via server. There should be no direct source to source link. This however, increases communication overhead in lieu of more reliable security feature.

It is also to be noted that we cannot fully trust the server. Server may attempt to get knowledge the knowledge of source's private data. This can be done in the following way.

Firstly, $c_1$ passes the information of its neighbor to the server for forwarding $R_1$ to other sources. If the server maliciously tries to capture the computed data from the first source node itself by declaring that $c_1$'s neighbors are already participated and none is left, the server asks $c_1$ to perform the SUM computation and sends that value to it. So, the data sends by $c_1$ is its own data itself. Server does this for $c_i$, i= 1,2, …, N, by subsequently choosing them as both initiator and the terminator node. In this way, server can get the private value of each of the source nodes. In order to avoid that, each time the initiator source node sends the SUM value to the server, it compares it with its private value, if both happen to be same; initiator source node sends the message to the server that "operation cannot be performed". Though, in this case a possibility of false alarm arises when the other sources' private values are all zeros. In a dense distributed system with large number of nodes, this possibility is very less.

It can be observed that to perform this operation aggregator need not be directly connected to all the nodes, but it has at least indirect connection to each node, i.e. source node 1 is connected to source node 2 and source node 2 is connected to source node N, which is directly connected to the server. Thus the server can reach source node 1 via source node N and source node 2. If less number of connections exists between the nodes then private data disclosure probability comes down (shown later) but key establishment becomes less secure due to unavailability of direct path [10].

## 6. Simulation Results

In this section, we show the performance of our scheme. In this work, we claim two distinct contributions over other privacy preservation data aggregation scheme like [1]. First is that in our scheme the probability of private data compromised is lesser and second is that our scheme has lesser time complexity. The second feature makes this scheme very much attractive for WSN, where the sensor nodes have very limited computational resources and processor capacity.

In CPDA [1] scheme, there exists certain probability where private data may be disclosed. This can only happen when the sink nodes exchange messages within the cluster. This can be estimated as

$$P(b) = \sum_{m=pc}^{Dmax} P(k=m)(1-(1-b^{m-1})^m)$$

Where *Dmax* = maximum cluster size, pc = minimum cluster size (= 3, two sink nodes and one aggregator), k = cluster size, b = probability that link level privacy is broken, P(k=m) = probability that a cluster size is m. In the case of our scheme, pc = Dmax = k = 2, $P(k=m)$ =1. So, we have plotted *P(b)* for CPDA and our scheme in fig. 2. It is observed that the probability of privacy compromised in CPDA has much steeper slope than that our proposed scheme.

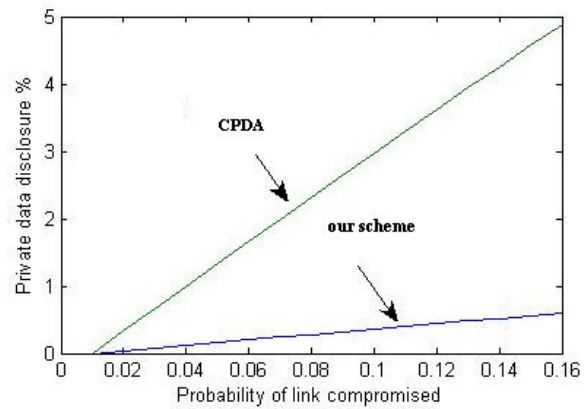

**Fig. 2 Probability of private data disclosure**

In fig. 2, we have compared the computational time requirement of our scheme to that of the CPDA scheme proposed in [1]. CPDA scheme and the corresponding algorithm leverage the algebraic properties of polynomials to compute the aggregated result. In CPDA algorithm, the computational time increases linearly with the addition of number of nodes. In our algorithm, we apply modular arithmetic, which incurs very less computational load and does have much effect with the addition of number of sources. It is to be noted that we have compared only the algorithm performance. As in CPDA, with number of client nodes increases, the computational time increases, we constraint number of source nodes to five. It is also impractical in CPDA to have large number of nodes in a single cluster. The comparison figure reveals the computational efficiency of our modular based algorithm. Our scheme has the additional advantage of the eliminating the complex cluster formation algorithm as in CPDA.

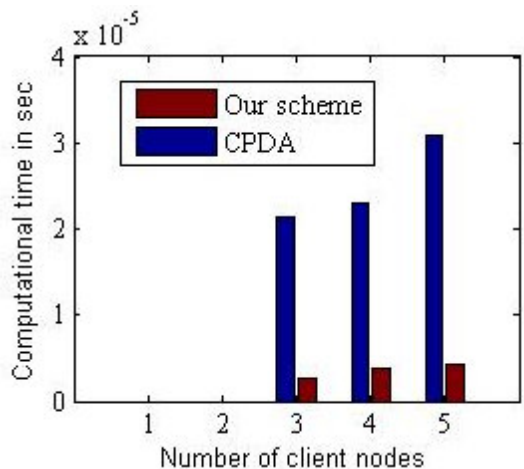

**Fig. 3 Computation time comparison between our scheme and CPDA [1]**

## 7. Conclusion

In this paper, we have presented a scheme for privacy preservation for secure data aggregation in WSN. In the proposed scheme, security of the data is ensured by a secured and robust key establishment policy. Privacy of the data of the source nodes is preserved by the concept of randomization through modular arithmetic. Our proposed scheme is basically a part of SMC concept [11]. We have shown the efficacy of our scheme in simulation results and also shown comparison with one existing scheme [1]. We have shown low computational time requirement for our scheme due to the simplicity of modular arithmetic as well as the lesser probability of private data disclosure due to secure key establishment. However, in this work we have not considered the communication overhead incurred in the overall scheme. This is included in our future work.

## 8. References

[1] W. He. et al, "PDA: Privacy-preserving Data Aggregation in Wireless Sensor Networks" IEEE INFOCOM, pp. 2045 – 2053, 2007.
[2] L.F. Cranor, (Ed.), "Special Issue on Internet Privacy," Comm. ACM, vol. 42, no. 2, Feb. 1999.
[3] A. Yao, "Protocols for secure computations," Proceedings of the 23rd Annual Symposium on Foundations of Computer Science, pp. 160-164, 1982.
[4] L. Sweeney, "Privacy Technologies for Homeland Security," Testimony before the Privacy and Integrity Advisory Committee of the Department of Homeland Security, Boston, MA, June 15, 2005.
[5] R. Agrawal and R. Srikant, "Privacy-Preserving Data Mining," ACM SIGMOD, 2000.
[6] Kargupta, H. et al. Random-data perturbation techniques and privacy-preserving data mining. Knowledge and Information Systems, vol.7, no.4, pp. 387-414, 2005.
[7] S. Goldwasser, "Multi-party computations: Past and present," 16th Annual ACM symposium on Principles of distributed computing, 1997.
[8] M. Conti, et al., "Privacy-preserving robust data aggregation in wireless sensor networks," Secur. Commun. Netw. vol. 2, pp. 195–213, 2009.
[9] M. Wright, et al., "Defending anonymous communications against passive logging attacks," IEEE Symposium on Security and Privacy, 2003.
[10] L. Eschenauer and V. D. Gligor, "A key-management scheme for distributed sensor networks," 9th ACM Conference on Computer and Communication Security, pp. 41–47, 2002.
[11] O. Goldreich, "Secure multi-party computation," Working Draft, Version 1.3, 2001.